# Picosecond energy transfer in quantum dot Langmuir-Blodgett nanoassemblies

Marc Achermann[a,*], Melissa A. Petruska[a], Scott A. Crooker[b], and Victor I. Klimov[a,*]

[a] *Chemistry Division, C-PCS, Los Alamos National Laboratory, Los Alamos, NM 87545*

[b] *National High Magnetic Field Lab, MST-NHMFL, Los Alamos National Laboratory, Los Alamos, NM 87545*

* *achermann@lanl.gov, klimov@lanl.gov*

We study spectrally resolved dynamics of Förster energy transfer in single monolayers and bilayers of semiconductor nanocrystal quantum dots assembled using Langmuir-Blodgett (LB) techniques. For a single monolayer, we observe a distribution of transfer times from ~50 ps to ~10 ns, which can be quantitatively modeled assuming that the energy transfer is dominated by interactions of a donor nanocrystal with acceptor nanocrystals from the first three "shells" surrounding the donor. We also detect an effective enhancement of the absorption cross section (up to a factor of 4) for larger nanocrystals on the "red" side of the size distribution, which results from strong, inter-dot electrostatic coupling in the LB film (the light-harvesting antenna effect). By assembling bilayers of nanocrystals of two different sizes, we are able to improve the donor-acceptor spectral overlap for engineered transfer in a specific ("vertical") direction. These bilayers show a fast, unidirectional energy flow with a time constant of ~120 ps.

## 1. Introduction

Non-radiative Förster energy transfer (ET) [1] via incoherent electrostatic interactions is an important communication and transport mechanism at the nanoscale. In living plants, this mechanism enables the delivery of energy from strongly absorbing chlorophyll antenna complexes to the reaction center; as a result the effective absorption cross section of this center can be greatly increased. The direction of energy flow in "natural" antenna complexes is controlled through the use of a series of coupled chromophores, which provide an energy gradient (an energy drop) toward the reaction center. Light-harvesting antenna functions have been reproduced in artificial energy-gradient assemblies based on organic molecules such as porphyrins (see, e.g., Ref. 2 and references therein). An alternative approach to engineered light-harvesting systems is the use of inorganic semiconductor nanocrystals [nanocrystal quantum dots (NQDs)]. Because of the effect of quantum confinement, the absorption and emission spectra of NQDs can be tuned over a wide energy range (~1 eV) by simply changing the particle dimensions. This size-controlled spectral tunability, combined with the ease with which NQDs can be manipulated into complex architectures, provides a powerful tool for engineering assemblies with controlled energy flows. ET in NQD systems has been studied using close packed solids of nominally monodisperse NQDs [3-5], random assemblies of NQDs with bi- [3, 4] and tri [5] -modal size distributions, and artificially engineered gradient structures [5]. The fastest ET time observed for NQDs has been ~1 ns, which is more than an order of magnitude slower than the limit (< 100 ps) estimated for the case of an NQD donor interacting with a single, resonant acceptor in its immediate vicinity [5].

Improvements in the ET rates are possible through the design of structures with reduced donor-acceptor separation and improved spectral overlap of the donor emission and acceptor absorption spectra. In this work, we demonstrate ET transfer times as fast as ~50 ps in close-packed, two-dimensional (2D) monolayers assembled using Langmuir Blodgett (LB) techniques. Based on the structural information obtained from transmission electron microscopy (TEM) studies, we are able to closely model complex, nonexponential ET dynamics within a *multishell* model. Furthermore, by using NQDs of two different sizes, we fabricate LB bilayers in which the donor-acceptor spectral overlap is engineered to provide efficient coupling of a monolayer of smaller NQDs to a larger NQD monolayer. These bilayers exhibit a fast directional ("vertical") ET with a time constant of ~120 ps.

## 2. Experimental

*Nanoparticle synthesis*

Hexadecylamine-capped CdSe NQDs were prepared as described in Ref. 6, and the NQD growth solution was washed with acetone and methanol to remove excess, unbound ligand. Ligand exchange of the hexadecylamine capping agent for octylamine was accomplished by stirring a solution of the NQDs in excess octylamine at 60 ºC for 12-24 hours. Hexane was added, and the NQDs were precipitated with methanol. TOPO/TOP-capped CdSe/ZnS core-shell NQDs were prepared as described in Ref. 7. Several precipitations of the NQDs with methanol were carried out to completely remove all unbound ligand.

*Langmuir-Blodgett films*

The Langmuir-Blodgett (LB) experiments were performed on a NIMA Technology (Coventry, England) Type 611 Langmuir trough (octylamine-capped CdSe NQDs) or with a KSV Instruments (Stratford, CT) 2000 system (CdSe/ZnS core-shell NQDs). A Barnstead NANOpure purification system produced water with a resistivity of 18 MΩ-cm for all experiments. Glass and quartz substrates were cleaned using the RCA procedure [8] and dried under a stream of nitrogen. The nanoparticles were dissolved in chloroform at concentrations between 2 and 10 mg/mL. The temperature of the subphase was 19.0 $\pm$ 1 ºC for all experiments. Surface pressure was measured with a paper Wilhelmy plate suspended from a NIMA or a KSV microbalance.

Pressure vs. area isotherms for the TOPO/TOP-capped CdSe/ZnS core-shell NQDs and the octylamine-capped CdSe NQDs were similar to those reported previously for analogous materials [9, 10]. In all cases, transfer pressures were chosen to achieve dense, close-packed films of NQDs. Single monolayers of TOPO/TOP-capped CdSe/ZnS core-shell NQDs were compressed at speeds of 15 mm/min to target pressures between 35 and 40 mN/m and transferred onto solid supports using conventional vertical dipping methods, in which the monolayer was transferred on the upward stroke onto hydrophilic substrates at rates between 2 and 5 mm/min. For the octylamine-capped CdSe NQDs, monolayers were compressed at similar speeds to a target pressure of 35 mN/m and transferred onto solid supports using the horizontal lifting method. To prepare the bilayer structures, the monolayer of large octylamine-capped CdSe NQDs was compressed and transferred onto a quartz substrate. This film was allowed to dry for up to 30 min before the monolayer of small octylamine-capped CdSe NQDs, compressed to 38 mN/m, was transferred on top using the horizontal lifting method. Monolayer films were transferred onto holey carbon-coated 200 mesh copper TEM grids for imaging. To prepare these samples, the sides of the grids were taped to glass substrates, and films were transferred onto these grids using the horizontal lifting method (octylamine-capped CdSe NQDs).

*Spectrally and time-resolved photoluminescence experiments*

To study the photoluminescence (PL) dynamics, NQD samples were excited at 3.1 eV by 100 fs pulses from a pulse-picked, frequency-doubled Ti:sapphire laser (Spectra-Physics Tsunami). The pump fluence was adjusted to excite less than 0.1 electron-hole pair per NQD on average. The use of low pump intensities allowed us to avoid inter- and intra-dot exciton-exciton interactions, which otherwise could complicate the interpretation of the experimental results. The PL of NQD samples was spectrally dispersed in a monochromator (Acton Research SpectraPro 300i with a 600 grooves/mm grating) and detected with a cooled multichannel plate photomultiplier tube (Hamamatsu R3809U-51). The detection system was coupled to a time-correlated single photon counting system (Becker-Hickl SPC-630), which allowed a 30-ps time resolution in time-resolved PL measurements. As indicated by previous studies [10], the PL parameters of NQD LB-films can change significantly within the first minutes of illumination. In order to avoid these slow transient changes, all PL data were acquired after sufficiently long, continuous illumination of the LB samples with the excitation laser beam to ensure that PL signals were stabilized. All measurements were performed at room temperature and under normal pressure.

## 3. Multishell model of ET in NQD LB assemblies

In NQD assemblies, ET occurs as a result of the electrostatic interactions of the emission dipole moment of an exciton generated in one NQD with the absorption dipole moment of another NQD. This process leads to the migration of excitons from smaller to larger NQDs in the ensemble, i.e., in the direction of the reduced energy gap. Exciton transfer is manifested as a low-energy shift of the PL band [3, 4] and a significant acceleration of the emission decay on the blue side of the spectrum [5]. As was observed in recent studies of ET dynamics [5], exciton migration is dominated by direct energy transfer from the “blue” to the “red” side of the PL spectrum across an energy gap of several tens of meV, which corresponds to the transition between NQDs that differ greatly in size. The latter observation is consistent with the energy structure of the band-edge exciton states shown in Fig. 1(a) [11, 12]. The band-edge optical properties of CdSe NQDs are governed by two manifolds of closely spaced levels. The lower exciton manifold responsible for the NQD emission has a relatively weak oscillator strength, whereas the upper manifold is characterized by a much stronger transition observed as an intense lowest (1S) absorption peak. The energy gap between the upper (absorbing) and lower (emitting) exciton manifolds is observed in optical spectra as a large (from ~20 to ~100 meV, depending on the NQD size [13]), global Stokes shift between absorption and emission peaks [Fig. 1(b)]. The exciton transfer between NQDs of similar sizes relies on coupling between two weak emitting transitions (the weak-coupling regime) and, therefore, is not efficient. Much stronger coupling is provided if the emitting transition in the donor NQD is resonant with a strong absorbing transition in the acceptor NQD (the strong-coupling regime). This situation can be realized in the case of NQDs with significantly different sizes. The ET between such NQDs results in the creation of a “hot” exciton in the acceptor NQD occupying the upper manifold of absorbing states. Because of the extremely efficient intraband relaxation (~1 ps or shorter [14]), the “hot” exciton rapidly loses its excess energy and relaxes to the acceptor ground state. This state produces emission, which is significantly red shifted with respect to the donor emission. Thus the strong-coupling mechanism results in a direct exciton transfer across a large energy interval that is roughly determined by the global Stokes shift.

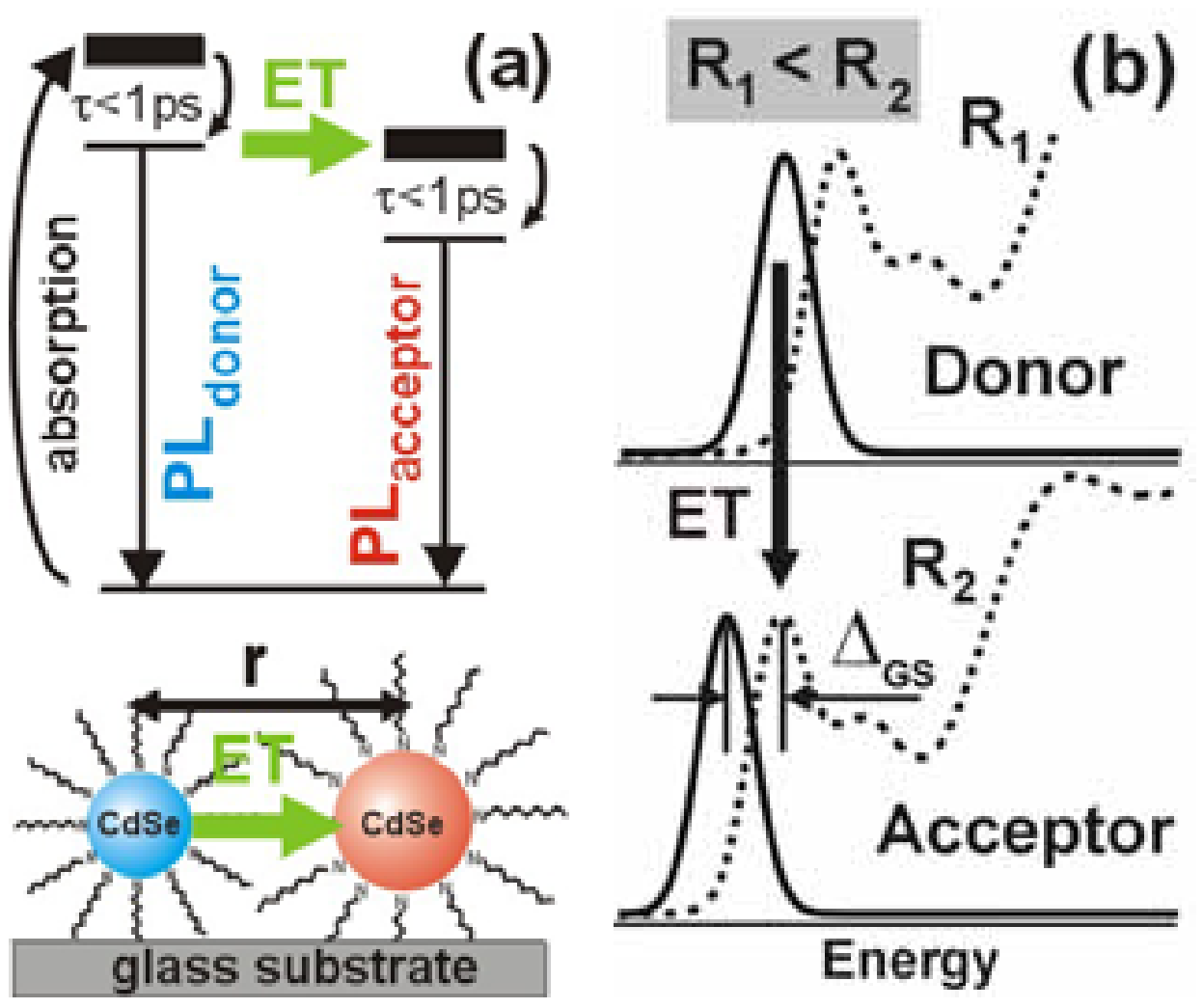

FIG 1: (a) Schematics of energy states in CdSe NQDs of two different sizes that interact via ET in the strong coupling regime (the emitting state in the smaller NQD is in resonance with the absorbing state in the larger NQD). The exciton generated in a smaller donor NQD by photoexcitation rapidly relaxes into its ground state. In the case of sufficiently close separation between the donor and acceptor NQDs, the exciton can experience resonant ET into the absorbing state of the acceptor NQD. After fast relaxation it recombines or is again transferred to another NQD in the assembly. (b) A schematic representation of the absorption (dashed line) and PL (solid line) spectra of donor and acceptor NQDs for the situation displayed in panel (a). The NQD radii ($R_1$ and $R_2$) are selected to provide a strong overlap between the emission spectrum in smaller NQDs and the lowest absorption peak in larger NQDs.

In the dipole-dipole approximation, the ET time, $\tau_{DA}$, between a donor and an acceptor NQD can be estimated using the Förster expression [1]

$$\frac{1}{\tau_{DA}} = \frac{2\pi}{\hbar}\frac{\mu_D^2\mu_A^2\kappa^2}{r^6 n^4}\Theta, \qquad \text{Eq. 1}$$

in which $\mu_D$ and $\mu_A$ are the donor and acceptor dipole moments, $r$ is the donor-acceptor separation, $\Theta$ is the overlap integral between normalized donor emission and acceptor absorption spectra, $\kappa^2$ is an orientational factor (accounts for the distribution of angles between the donor and the acceptor dipole moments; for random dipole orientation $\kappa^2 = 2/3$), and $n$ is the refractive index of the medium. Most efficient ET corresponds to the situation for which two NQDs located in the immediate vicinity of each other resonantly interact in the strong-coupling regime. It has been determined that for CdSe NQDs of ~ 2.2 nm radius (R) with a surface-to-surface separation of ~1 nm (the typical length of a surface ligand molecule), the ET time is fast (~38 ps) if the “emitting” transition in a donor NQD is resonant with an “absorbing” transition in an acceptor NQD [5] (spectral overlap $\Theta$ is optimized). However, in films fabricated by drop casting from nominally monodisperse NQDs, the ET times found experimentally are on the nanosecond time scale. As was concluded in Ref. 5, this relatively slow transfer does not result from weak inter-dot coupling but from a low NQD packing density. The small number of NQDs in the immediate vicinity of a donor NQD (in the first shell) reduces significantly the probability of the occurrence of a resonant acceptor. Therefore, in drop-cast films, ET is dominated by interactions of a donor NQD with the *second* or *higher-order* shells of nearest neighbors, which have a higher NQD population than the first shell and hence provide a greater probability for the occurrence of the resonant acceptor.

Because of the enhanced NQD packing density induced by monolayer compression, LB films can, in principle, allow an enhanced efficiency of Förster interactions between NQDs at the nearest-neighbor distance by providing an increased number of potential acceptors in the first shell. Furthermore, an additional enhancement compared to three-dimensional (3D) films can occur because of the decreased screening of inter-dot Coulomb interactions, which takes place not through a high index NQD solid, which is the case in 3D films, but rather through a lower index medium (substrate on one side of the LB monolayer and vacuum or air on the other side).

A typical TEM image of an LB film in Fig. 2(a) indicates that the NQDs form a dense, close-packed monolayer, which exhibits short-range, hexagonal order. In the case of an ideal hexagonal 2D array [Fig. 2(b)], NQDs surrounding a specific “center” dot can be grouped into shells according to their distance from the central dot. Furthermore, shells of similar radii (sub-shells; shown in Fig. 2(b) by different shadings) can be grouped into main shells (shown in Fig. 2(b) by different colors).

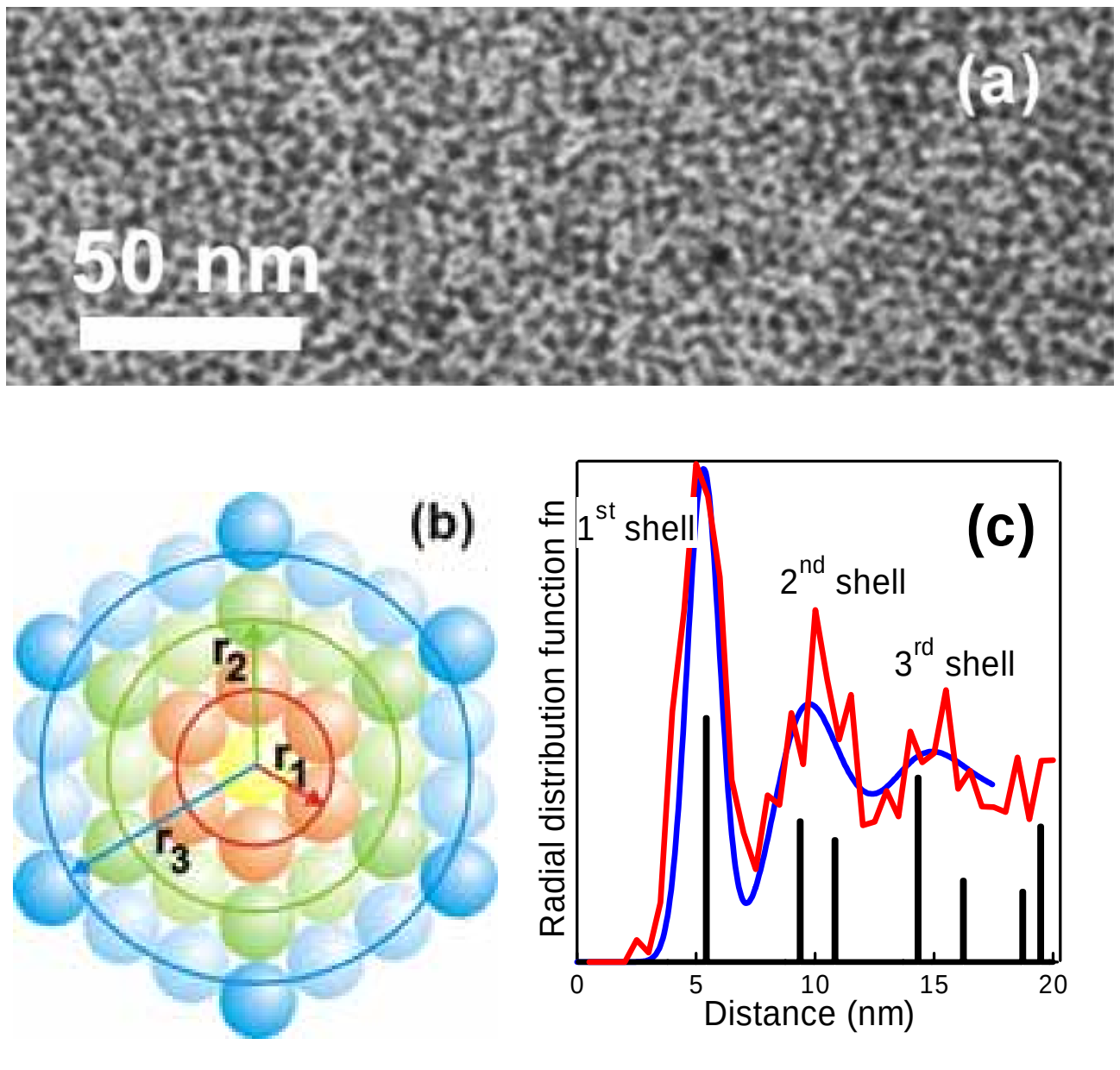

FIG 2: (a) A TEM image of an LB monolayer of octylamine-capped CdSe NQDs with radius $R$ = 2.2 nm (the monolayer was transferred onto a TEM grid at a pressure of 35 mN/m). (b) An ideal hexagonal 2D array of spheres. Different colors shows three main shells. Different shadings within the main shells show different sub-shells. (c) RDF of the LB sample shown in panel (a) (red line). Black bars represent the RDF of an ideal hexagonal array composed of spheres with a center-to-center distance of 5.4 nm. The blue line is the RDF of a “nonideal” hexagonal array with fluctuation in the sphere-to-sphere distances of 13%.

These main shells have radii that are approximate multiples of the nearest-neighbor distance, d: $r_1/d = 1$, $r_2/d = 1.87$ and $r_3/d = 2.76$ ($r_i$ is the mean radius of sub-shells that compose the main shell number *i*). The packing of the film can be characterized by the radial distribution function (RDF). The ideal hexagonal array has a discrete RDF (shown in Fig. 2(c) for d = 5.4 nm), in which each δ-function-like peak represents a distinct shell. For our LB monolayers, we determine the RDF from the distribution of inter-dot distances in TEM images. As a consequence of the short-range order, the RDF of LB samples exhibits several maxima at small *r*. For our samples, we typically observe three distinct RDF peaks [Fig. 2(c)], indicating the existence of three well-defined shells. The measured distribution function can be closely reproduced using the convolution of the RDF of an ideal hexagonal lattice [discrete bars in Fig. 2(c)] with a Gaussian distribution function (standard deviation of 13%), which accounts for the distribution of the inter-dot distances (note, that this distribution is not identical to the NQD size distribution and is typically broader than this value by a factor of ~2). The population factors for individual shells ($N_j$; $j$ = 1, 2, …) can be quantified by integrating the individual peaks of the RDF. By applying this procedure, we find that the first three shells contain ca. 5, 8, and 11 NQDs. These values are lower than those characteristic of the ideal hexagonal array ($N_1 = 6$, $N_2 = 12$, and $N_3 = 18$), indicating that there is room for improvement in the packing density.

The existence of well-defined shells in LB films can simplify the analysis of the ET dynamics. Specifically, instead of considering ET in terms of NQD-to-NQD interactions, we will consider it in terms of NQD-to-shell interactions. The efficiency of the latter process is determined by the shell radius ($r_i$) and its population factor ($N_i$), which both can be derived from the TEM analysis. For a given shell number *i*, the effective transfer rate ($\tau_i^{-1}$) can be expressed as $\tau_i^{-1} \propto N_i r_i^{-6}$.

As a measure of ET dynamics we use a time-dependent average exciton energy $\langle E(t) \rangle$. For a given donor NQD (emission energy $E_D$) interacting with the *i*th-shell, the temporal evolution of the exciton energy can be described as $E(t) = E_A + (E_D - E_A)e^{-\frac{t}{\tau_i}}$, in which $E_A$ is the acceptor emission energy. According to Ref. 5, in assemblies of nominally monodisperse NQDs, ET is dominated by the strong-coupling mechanism for which the exciton is transferred across a large energy interval ($\Delta = E_D - E_A$), which is roughly determined by the global Stokes shift. In this case the time dependent exciton energy can be presented as $E(t) = E_A + \Delta e^{-\frac{t}{\tau_i}}$. In order to extend this single-shell expression to a multi-shell situation existing in real LB assemblies, we can simply average it over all shells using weighting factors, $f_i$ ($\sum_{i=1}^{\infty} f_i = 1$), that describe the probability for an exciton in a given donor NQD to undergo transfer into the shell number *i*. This averaging leads to the expression

$$\langle E(t) \rangle = \langle E_D \rangle - \Delta \sum_{i=1}^{\infty} f_i (1 - e^{-\frac{t}{\tau_i}}) = \langle E_A \rangle + \Delta \sum_{i=1}^{\infty} f_i e^{-\frac{t}{\tau_i}}, \text{ Eq. 2}$$

in which $\langle E_D \rangle$ and $\langle E_A \rangle$ are the average energies of excitons in donor and acceptor NQDs, respectively. The latter expression indicates that the exciton dynamics within the shell model can be described by the sum of exponential decay terms, and each term is directly associated with ET into a specific shell.

## 4. ET dynamics in LB monolayers

Figure 3 (a) displays spectrally resolved PL dynamics of an LB monolayer fabricated from CdSe/ZnS core-shell NQDs with a core radius of 1.5 nm. The PL traces show a progressively faster decay with increasing spectral detection energy. This behavior is a clear signature of ET that leads to the migration of excitons from smaller to larger NQDs in the NQD ensemble [5]. The observed dynamics, however, are not entirely due to exciton inter-dot transfer but also result from radiative and nonradiative recombination. In order to extract the contribution related purely to ET, we normalize the recorded PL traces by a time dependent factor, $\beta(t) = \int_0^{\infty} I(\omega, t) d\omega$, in which $I(\omega,t)$ is the time- and spectrally resolved PL intensity. This factor is proportional to the total number of excitons at a given time $t$ and describes the exciton recombination dynamics from both radiative and nonradiative processes [15]. The normalized traces [Fig. 3(b)] indicate that the signal decay on the blue side of the PL spectrum is complementary to PL growth on its red side, which corresponds to the outflow (in smaller NQDs) and inflow (in larger NQDs) of excitons, respectively. ET also leads to a pronounced dynamic red shift in the PL emission spectrum; it moves by 60 meV within 10 ns [Fig. 3(c)].

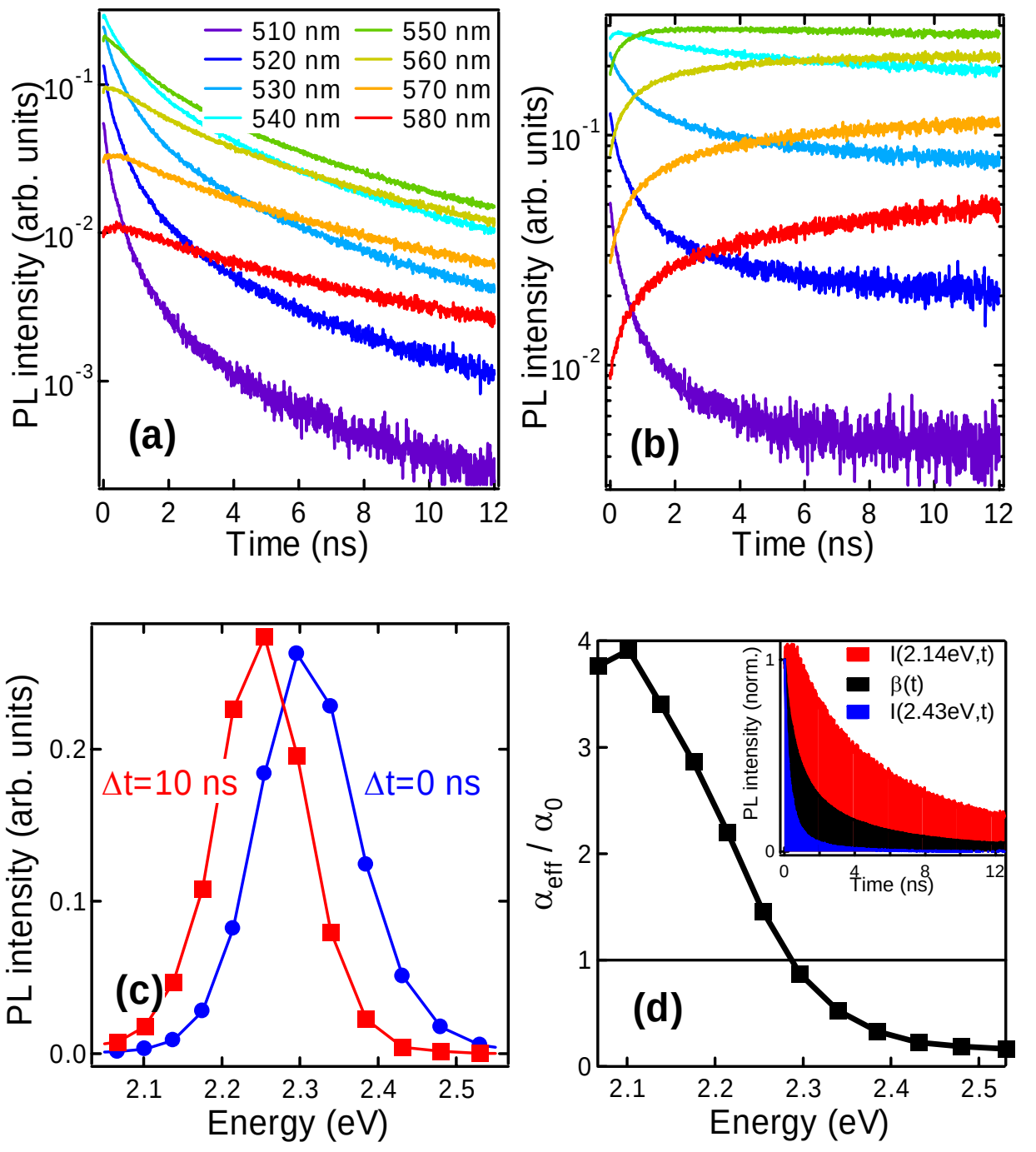

FIG 3: (a) Spectrally resolved PL dynamics in an LB monolayer of core-shell CdSe/ZnS NQDs (the CdSe core radius is 1.5 nm). (b) The same dynamics but corrected for recombination losses; these dynamics are primarily due to ET. (c) PL spectra measured at 0 ns (blue solid circles) and 10 ns (red squares) after excitation indicating a large red shift resulting from ET. (d) Relative effective absorption cross-section $\sigma_{ET}/\sigma_0$ plotted as a function of the emission energy. The inset shows *β(t)* (black) and PL traces [$I(\omega,t)$] at 2.14 eV (red) and 2.43 eV (blue), all normalized to zero time delay. The areas under these traces are used to calculate the spectrally dependent effective absorption cross section.

The ET-induced increase in the PL intensity on the red side of the spectrum can be interpreted in terms of an effective increase of the absorption cross section in larger NQDs due to the exciton inflow from smaller NQDs (the latter perform "light-harvesting" antenna functions). We define the effective ET-absorption cross section $\sigma_{ET}(\omega)$ of NQDs with a given emission energy, $\omega$ (i.e., a given radius $r$), as the product of the usual absorption cross section without ET, $\sigma_0$, and an ET-induced enhancement factor, which is given by the ratio between the numbers of photons emitted by a given NQD with ET, $I(\omega,0)^{-1}\int_0^\infty I(\omega,t)dt$, and without ET, $\beta(0)^{-1}\int_0^\infty \beta(t)dt$ [see the inset of Fig. 3(d)].

Figure 3(d) displays the ratio $\sigma_{ET}(\omega)/\sigma_0$ as a function of the NQD emission energy. These data indicate a considerable increase (by a factor of ~4) in the effective absorption cross section in larger NQDs (the red side of the PL band), which is accompanied by a reduction in the effective absorption cross section in smaller NQDs (the blue side of the emission spectrum).

ZnS-overcoated CdSe NQDs give rise to very stable, highly luminescent LB films. However, the ZnS shell increases the overall NQD size and hence the distance between adjacent NQDs. Because of the $r^{-6}$ distance dependence of the ET rate, even a small increase in the donor-acceptor separation results in considerably decreased ET rates. In order to achieve faster ET we use "bare" NQDs (no ZnS shell) capped with relatively short ligand molecules. We perform spectrally and time-resolved PL measurements of an LB-film of octylamine-capped CdSe NQDs with radius $R$ = 2.2 nm. From the measured PL traces we derive the temporal evolution of the average exciton energy using the expression:

$$\langle E(t)\rangle = \beta^{-1}(t)\int_0^\infty \hbar\omega\cdot I(\omega,t)d\omega . \quad \text{Eq. 3}$$

The resulting $\langle E(t)\rangle$ dynamics is shown in Fig. 4 by solid circles. To evaluate the role of increased packing density (presumably achieved by the LB technique) on ET rates, we compare these dynamics with those measured for drop-cast (3D) films fabricated from the same NQDs (triangles). The recorded traces indicate that the ET-induced shift in the average exciton energy is significantly greater for the LB film (55 meV) compared to the drop-cast sample (30 meV). Furthermore, we observe that the initial ET dynamics are significantly faster in the LB monolayer. The initial ET time constant is 50 ps, which is as fast as in, e.g., porphyrin-based artificial antenna complexes [2]. These fast dynamics are not present in the drop-cast sample, for which the fastest transfer time is ~850 ps. For two adjacent NQDs of 2.2 nm radius with a surface-to-surface distance of 1 nm (the length of the ligand molecule), the estimated ET time is approximately 40 ps for the case of the strong-coupling regime (the dipole moments and the spectral overlap integral used in this estimation are from Ref. 5). This time is very close to the value observed in the LB film suggesting that in LB monolayers, there is a significant probability for the occurrence of a resonant acceptor in the first shell of the nearest neighbors. In contrast, this probability is quite low in the less dense drop-cast films for which the fastest ET time is on the order of a nanosecond.

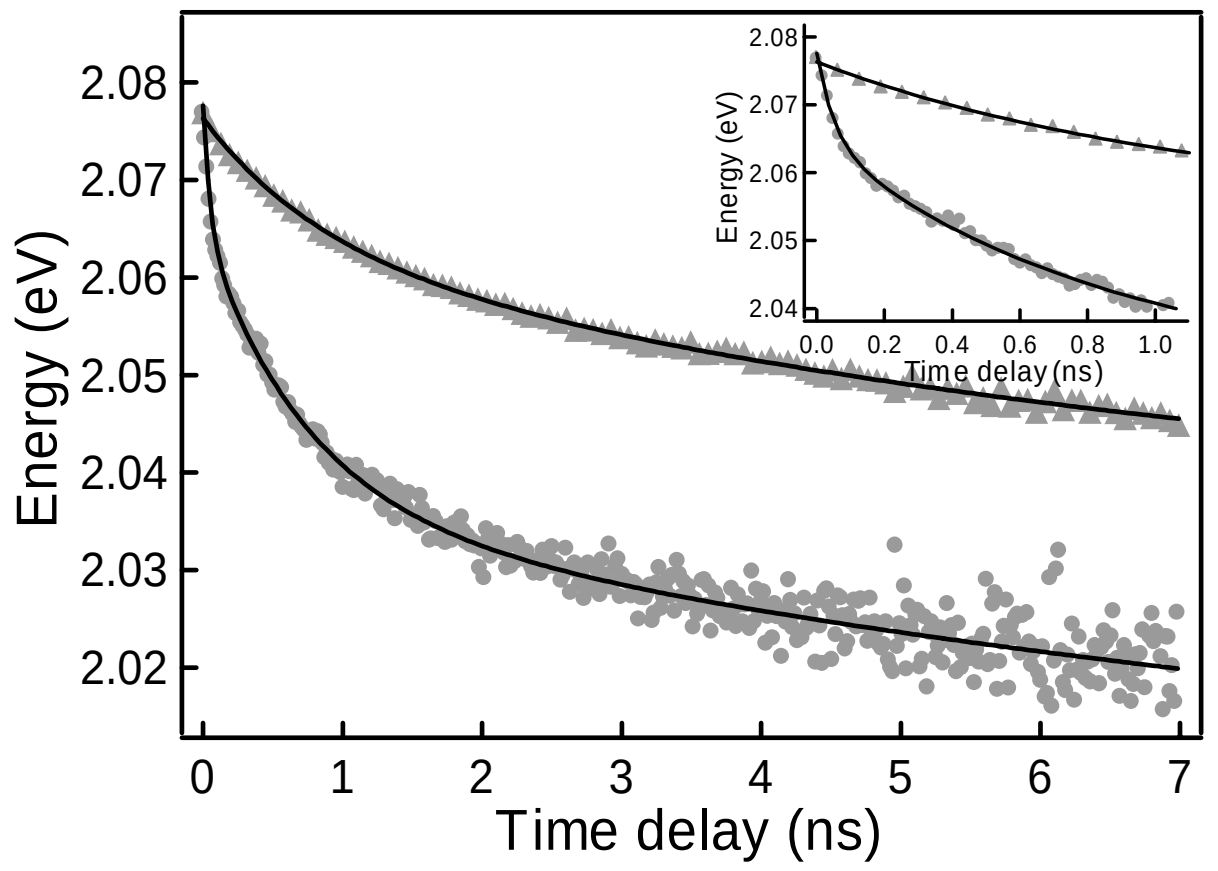

FIG 4: (a) Comparison of the dynamics of the average exciton energy $\langle E(t)\rangle$ in an LB monolayer (circles) and a drop-cast film (triangles) fabricated from the same NQDs (R = 2.2 nm, octylamine capping). These dynamics can be accurately fit to either a two- (drop-cast film) or three-exponential (LB monolayer) decay. Inset shows the same dynamics on a shorter time scale.

In order to apply the shell model to the analysis of ET dynamics in LB assemblies, we first estimate the range of Förster transfer (in terms of shell numbers) by comparing its efficiency with the efficiency of the intrinsic radiative decay. The critical shell number ($i_F$) for which ET and radiative recombination are equally efficient can be estimated from the condition $\tau_r = \tau_{i_F}$, in which $\tau_r$ is the radiative decay time. The ET time for the $i$th shell scales with respect to the first-shell transfer time as $\tau_i = i^6\tau_1\frac{N_1}{N_i}$. Since in a 2D monolayer the shell occupation factor, $N_i$, is approximately proportional to $i$, the expression for $\tau_i$ reduces to $\tau_i \approx i^5\tau_1$, which ultimately gives $i_F \approx \sqrt[5]{\frac{\tau_r}{\tau_1}}$. For the CdSe NQD sizes used in our experiments, the room-temperature radiative decay time is ~20 ns [16], while the measured $\tau_1$ time is ~50 ps, which yields $i_F \approx 3$. This result indicates that in the analysis of ET dynamics in our LB sample, it is sufficient to account for only the three first shells interacting with the donor NQD. In this case, we should be able to accurately fit the ET dynamics to a sum of three exponential terms. We indeed find that such a fit is possible (see Fig. 4), and it produces time constants $\tau_1$ = 50 ps, $\tau_2$ = 0.75, ns and $\tau_3$ = 10 ns. Each of these time constants can presumably be assigned to a specific shell and, therefore, should be directly related to the shell radius and its occupation factor. Specifically, the Förster theory [Eq.(1)] predicts that the shell radii and corresponding transfer time constants are related by the expression: $\frac{r_i}{r_j} = \sqrt[6]{\frac{N_i}{N_j}\frac{\tau_i}{\tau_j}}$. Using the time constants determined by fitting the experimental data in Fig. 4 and the shell occupation factors derived from the TEM analysis (section 3), we find $r_2/r_1$ = 1.7 and $r_3/r_1$ = 2.8. Both of these values are in good agreement with those expected for the ideal hexagonal array (1.87 and 2.76), indicating that time constants derived from a simple multiexponential analysis are indeed directly associated with ET into specific NQD shells.

## 5. Energy transfer in LB-bilayers

ET in LB monolayers built of nominally monodisperse NQDs relies on the accidental occurrence of a spectral overlap in donor-acceptor pairs with a sufficiently close spatial separation. Therefore, in ET dynamics of these monolayers we observe time constants related to interactions not only between adjacent NQDs (the first shell ET) but also between NQDs at relatively large distances in the case for which a viable acceptor is not found in the first shell. One can promote the nearest-neighbors interactions by intentionally placing an acceptor, which provides a good spectral overlap, in the immediate vicinity of a donor NQD. One example of such a structure is an NQD bilayer built of NQDs with distinctly different sizes [5]. In the case of a large size disparity, an exciton in a smaller NQD strongly couples to a large-spectral density manifold of high energy states in a larger NQD, which can lead to efficient inter-layer ET. Such a prototype bilayer structure was realized in Ref. 5, where covalent linkages were used between NQDs of 1.3 and 2.05 nm radii. An efficient directional ("vertical") ET was observed. However, because of the relatively large length of the linker molecule, which caused a large inter-dot separation (~6.2 nm center-to-center), the "vertical" ET was relatively slow and had a time constant of ~750 ps.

In order, to improve inter-layer coupling and to achieve a faster "vertical" ET, we fabricate a bilayer structure [Fig. 5(a)] by directly transferring an LB monolayer of small NQDs (1.5 nm radius) onto an LB monolayer of larger NQDs (1.9 nm radius). This method eliminates the linker and allows us to minimize the distance between donor and acceptor NQDs, which is determined by the length of the surface ligand molecule (~1 nm). In figure 5(b), we display the spectra of this structure recorded at different times after the excitation. The spectra are normalized to have equal spectral area, which allows us to visualize dynamics that are predominantly due to ET (we assume that both radiative and non-radiative decay rates are independent of NQD size, and hence, of spectral energy [15]). At zero time delay, the measured spectra can be decomposed into two PL bands of similar amplitudes [approximated by two Gaussian bands in Fig. 5(b)] associated with emission from individual monolayers. With increasing delay time, the high energy PL (smaller NQD layer) rapidly decays, while the low-energy PL (larger NQD monolayer) grows in. These observations are consistent with the "vertical" inter-layer ET in which excitons migrate from smaller to larger NQDs. In addition to the gradual increase in the amplitude, the low-energy PL band shows a red shift, which is due to "horizontal" ET within the monolayer of larger NQDs, as discussed in the previous section.

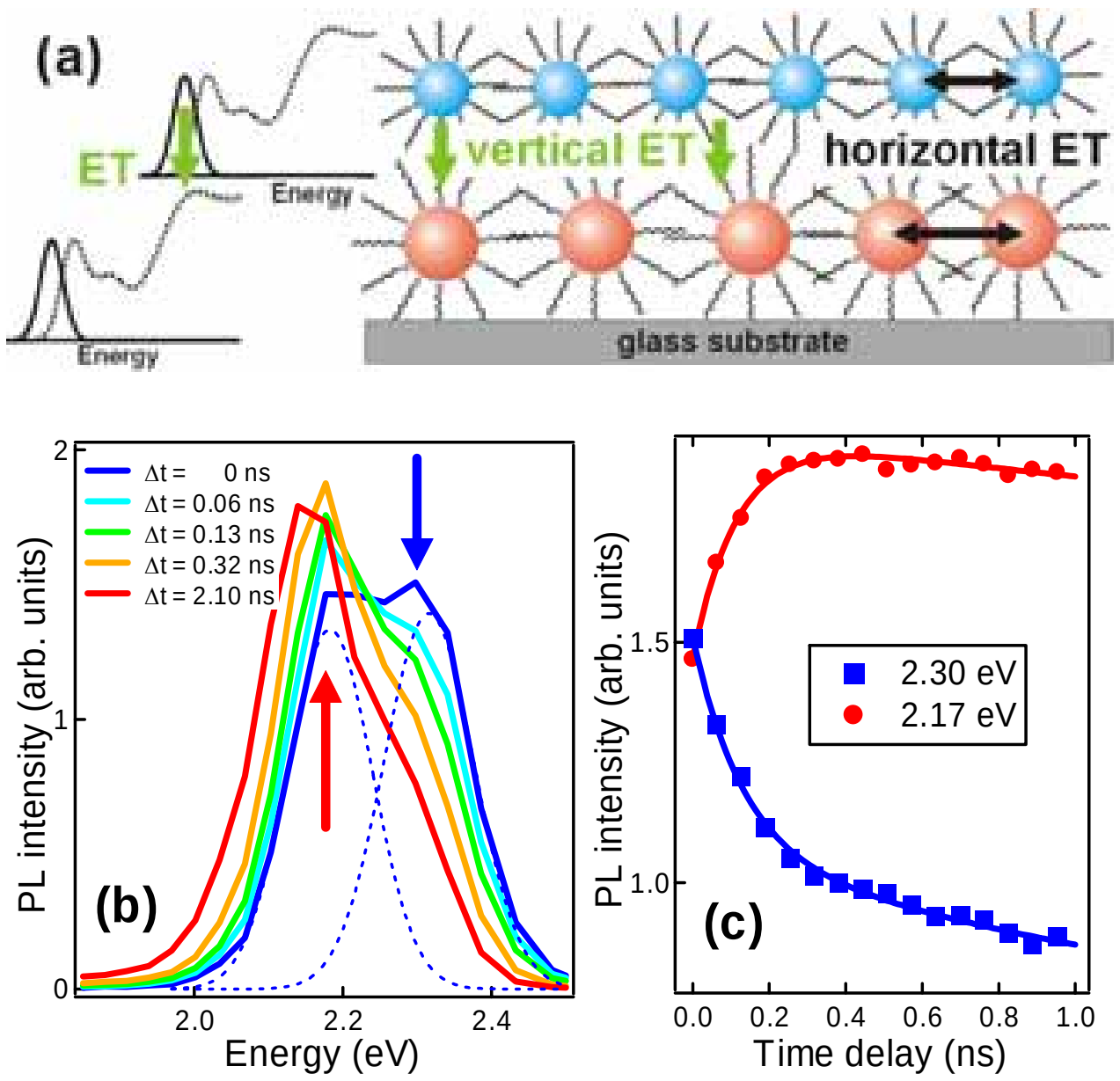

FIG 5: (a) Schematic of the "energy-gradient" bilayer structure fabricated from NQDs with a significant size disparity (panel on the right). This design allows us to strongly couple the emission from the layer of smaller NQDs to the dense, strongly absorbing manifold of high-energy states in larger NQDs (panel on the left). (b) Normalized PL spectra of the NQD bilayer measured at different times after excitation. The zero-delay-time spectrum is decomposed into two Gaussian bands (blue dashed lines) that correspond to emission spectra from individual monolayers. (b) The bilayer PL dynamics measured at 2.30 eV (blue squares) and 2.14 eV (red solid circles). The use of these spectral energies [indicated by arrows in panel (b)] allows us to monitor dynamics related primarily to "vertical" inter-layer transfer.

In Fig. 5(c), we directly compare ET-related dynamics in the two layers by displaying normalized PL traces recorded at 2.30 eV and 2.17 eV, which approximately correspond to the centers of the zero-time PL bands in monolayers of the smaller and larger NQDs, respectively. As our studies of single monolayers (Section 4) indicate, the PL changes at these spectral energies induced by "horizontal" energy transfer are not significant (exciton inflow and outflow approximately compensate each other); therefore, it is convenient to use these energies for studies of "vertical" transfer. The complementary short-term dynamics observed at 2.30 eV and 2.17 eV are consistent with the assumption that they are indeed primarily due to "vertical" ET. The exponential fit to the initial PL decay at 2.30 eV and its growth at 2.17 eV yields that same time constant of ~120 ps, which characterizes the inter-layer transfer. Although the ET is faster than in previously assembled bilayer structures, the observed time is still slower than the time constant derived for the first-shell transfer in LB monolayers. The fact that the "vertical" transfer is slower than the "horizontal" one is likely because the inter-dot separation within the compressed layer is smaller than the dot-to-dot separation in the vertical direction, along which no pressure was applied.

## 6. Summary

We have studied spectrally resolved ET dynamics in monolayers and bilayers of CdSe NQDs fabricated using LB techniques. We observe that in compressed LB monolayers, the ET time constant can be as fast as 50 ps, which is close to the theoretical limit for the exciton transfer between two adjacent NQDs resonantly interacting in the strong-coupling regime. Furthermore, by analyzing spectrally resolved PL dynamics, we find that ET is dominated by interactions of a center donor NQD with NQD acceptors from the first three shells; as a result the ET-related PL dynamics can be fit by a sum of three exponential decay terms in which each term is associated with a specific shell. The NQD-to-shell transfer time constant can be described by the Förster theory, in which we use the shell radii and population factors extracted from the TEM analysis of the sample morphology. A consequence of fast ET in NQD LB films is the considerable enhancement (up to a factor of 4) of the effective absorption cross-section for larger NQDs in the NQD ensemble (the light harvesting effect).

To promote nearest-neighbor interactions, we fabricate LB bilayers built of NQDs with distinctly different sizes ("energy-gradient" structures). In these structures, the excitons in a layer of smaller NQDs strongly couple to a dense manifold of high-energy states in the larger NQDs comprising the second layer. This coupling results in an efficient, inter-layer, "vertical" transfer with a time constant of ~120 ps. Considering that such vertical transfer involves an energy step on the order of the global Stokes shift, we could envision an energy-gradient structure with ~20 layers of CdSe NQDs that can efficiently collect energy across the entire visible range and directionally transport this energy over a 100 nanometer distance within a few nanoseconds (much faster than the radiative decay). This large range of exciton transport, combined with the ability to tailor the absorption spectra of NQDs, provides an interesting opportunity for the development of NQD-based light-harvesting structures that can be applied, e.g., in artificial photosynthesis or photovoltaic devices.

This work was supported by Los Alamos Directed Research and Development Funds, and the U. S. Department of Energy, Office of Sciences, Division of Chemical Sciences.